\documentclass[conference]{IEEEtran}
\IEEEoverridecommandlockouts
\usepackage{cite}
\usepackage{amsmath,amssymb,amsfonts}
\usepackage{algorithmic}
\usepackage{graphicx}
\usepackage{textcomp}
\usepackage{xcolor}

\usepackage{amsmath,amsfonts}
\usepackage{algorithmic}
\usepackage{algorithm}
\usepackage{array}
\usepackage[caption=false,font=normalsize,labelfont=sf,textfont=sf]{subfig}
\usepackage{textcomp}
\usepackage{stfloats}
\usepackage{url}
\usepackage{verbatim}
\usepackage{graphicx}
\usepackage{listings}
\usepackage{cite}
\usepackage{tikz}
\usepackage{pgfplots}
\usepackage{pgfplotstable}
\pgfplotsset{compat=1.18}

\usepackage[top=0.75in, bottom=1.1in, left=0.673in, right=0.673in]{geometry}

\usepackage[T1]{fontenc}    
\usepackage{textcomp}       
\usepackage{upquote}        
\usepackage{xcolor}
\usepackage{caption}

\def\BibTeX{{\rm B\kern-.05em{\sc i\kern-.025em b}\kern-.08em
    T\kern-.1667em\lower.7ex\hbox{E}\kern-.125emX}}

\begin{document}

\title{A Comparative Analysis of zk-SNARKs and zk-STARKs: Theory and Practice
}

\author{\IEEEauthorblockN{
Ayush Nainwal,
Atharva Kamble and 
Nitin Awathare }

\IEEEauthorblockA{Department of Computer Science and Engineering, Indian Institute of Technology, Jodhpur}
\IEEEauthorblockA{\{iayush.n2, atharvakamble028\}@gmail.com,\{nitina\}@cse.iitb.ac.in, \{nitina\}@iitj.ac.in}}

\maketitle

\begin{abstract}
Zero-knowledge proofs (ZKPs) are central to secure and privacy-preserving computation, with zk-SNARKs and zk-STARKs emerging as leading frameworks offering distinct trade-offs in efficiency, scalability, and trust assumptions. While their theoretical foundations are well studied, practical performance under real-world conditions remains less understood.

In this work, we present a systematic, implementation-level comparison of zk-SNARKs (Groth16) and zk-STARKs using publicly available reference implementations on a consumer-grade ARM platform. Our empirical evaluation covers proof generation time, verification latency, proof size, and CPU profiling. Results show that zk-SNARKs generate proofs 68x faster with 123x smaller proof size, but verify slower and require trusted setup, whereas zk-STARKs, despite larger proofs and slower generation, verify faster and remain transparent and post-quantum secure. Profiling further identifies distinct computational bottlenecks across the two systems, underscoring how execution models and implementation details significantly affect real-world performance. These findings provide actionable insights for developers, protocol designers, and researchers in selecting and optimizing proof systems for applications such as privacy-preserving transactions, verifiable computation, and scalable rollups.
\end{abstract}


\section{Introduction}
In the rapidly advancing field of cryptography, zero-knowledge proof systems (ZKPs) have become fundamental for ensuring secure and privacy-preserving computations. These systems enable a prover to demonstrate the validity of a statement to a verifier without disclosing any information beyond the statement's truth. 
For instance, ZKPs are integral to secure blockchain transactions, where they allow users to prove ownership or correctness of transactions without exposing transaction details \cite{6956581} \cite{buterin2021zkSnarksIntro}. Similarly, in identity verification, ZKPs enable users to authenticate themselves without revealing unnecessary personal data \cite{buterin2024binius} 
They are also crucial in privacy-preserving computations, allowing complex data operations to be verified securely without compromising the underlying data \cite{cryptoeprint:2016/116}.

Among the various frameworks developed for zero-knowledge proofs, zk-SNARKs (Zero-Knowledge Succinct Non-Interactive Arguments of Knowledge) and zk-STARKs (Zero-Knowledge Scalable Transparent Arguments of Knowledge) stand out as two leading technologies. However, each comes with its strengths and trade-offs. 


zk-SNARKs are highly efficient due to their compact proof sizes and fast verification times, making them suitable for 
constrained environments such as blockchain where longer verification time would raise the significant security vulnerabilities, such as reduce the mining power utilization \cite{info15080463}.

zk-STARKs, on the other hand, eliminate the need for a trusted setup, making them inherently more secure and transparent \cite{SassonBHR18}. They leverage transparent computation methods that do not rely on pre-generated parameters, enhancing trust and removing the risks  associated with zk-SNARKs' trusted setup \cite{buterin2017starksPart1}. Additionally, zk-STARKs are post-quantum secure, meaning they can resist attacks from quantum computers, a feature that is increasingly important as quantum technologies advance \cite{buterin2018starksPart3}. However, this added security comes at a cost: zk-STARKs have larger proof sizes and require more computational resources, which can limit their applicability in scenarios where efficiency is paramount, specifically in terms of block validation time~\cite{buterin2024binius}.

A systematic empirical comparison of zk-SNARKs and zk-STARKs is crucial, as their trade-offs in trust, scalability, efficiency, and post-quantum security often diverge in practice due to implementation and real-world constraints.


Prior works, such as Buterin~\cite{buterin2016qapHero} and Ben-Sasson et al.~\cite{cryptoeprint:2013/879}, have laid the theoretical foundations of zk-SNARKs and zk-STARKs, exploring their cryptographic assumptions, algebraic constructions, and asymptotic complexities. These studies provide deep mathematical insights into the security and scalability trade-offs of both proof systems. However, they often stop short of quantifying how these protocols behave in practice, leaving open questions around implementation bottlenecks, proof sizes, and the real costs of verification in diverse computational settings.

This paper addresses that gap by contributing a systematic, implementation-level comparison of zk-SNARKs and zk-STARKs. We empirically evaluate proof generation times, verification latencies, proof sizes, and CPU profiles, revealing how practical performance can diverge from theoretical expectations due to factors such as execution models, cryptographic libraries, and on-chain constraints. By grounding the trade-offs in concrete data, our study provides actionable insights for developers, protocol designers, and researchers—informing the choice of proof system for privacy-preserving transactions, verifiable computation, and scalable rollups—while also guiding future optimization of proving systems \cite{Samardzic2024Accelerating}.


We have evaluated both the protocols for empirical comparison on the experimental setup lies in its use of a consumer-grade ARM-based laptop (Apple M1). Our analysis uses publicly available reference implementations of Groth16 and STARKs
, providing transparent baselines for proof generation and verification and allowing us to focus on comparative measurement over implementation-specific optimizations\footnote{The Groth16 and STARK implementation respectively, are as follows \\ \url{https://github.com/arnaucube/go-snark-study}\\ \url{https://github.com/actuallyachraf/zkstarks}}.


Key findings reveal zk-SNARKs proved 68× faster to generate proof (55 ms vs. 3809 ms) but slower to verify (1807 ms vs. 472 ms) than zk-STARKs, and required a trusted setup. Conversely, zk-STARKs produced larger proofs (69 KB vs. 0.6 KB) but verified faster without trusted setup. CPU profiling showed zk-SNARKs bottlenecked by pairing and field arithmetic, while zk-STARKs were limited by polynomial and big-integer operations, highlighting how implementation choices heavily influence real-world performance.

The remainder of this paper is organized as follows: \S\ref{sec:pract} presents the theoretical comparison of zk-SNARKs and zk-STARKs, serving as a bridge to the empirical analysis. \S\ref{sec:practicalImplementation} outlines their practical implementations, while \S\ref{sec:evaluation} details the experimental setup and key findings from our empirical study. Related work is reviewed in \S\ref{sec:related}, and we conclude with discussion and future directions in \S\ref{sec:discussion}.

\section{Theoretical Comparison of zk-SNARKs and zk-STARKs}
\label{sec:pract}

This section bridges the theoretical framework and empirical analysis, providing context for the practical results presented later. We begin with the mathematical foundations, covering computation representation, proof construction, and security assumptions, followed by a discussion of efficiency and scalability, including proof size, verification speed, and empirical measurement of prover-side overheads.





\subsection{Mathematical Foundations}
This section compares zk-SNARKs and zk-STARKs in terms of computation representation, proof construction, and security assumptions.

\subsubsection{Representation of Computation}

Both zk-SNARKs and zk-STARKs convert computations into algebraic forms to enable succinct, verifiable proofs, but the representations differ fundamentally. 
zk-SNARKs use the \textit{Rank-1 Constraint System (R1CS)}~\cite{9315064, cryptoeprint:2016/260}, encoding constraints \(A \cdot B = C\) over input and witness variables. These constraints are then translated into \textit{Quadratic Arithmetic Programs (QAPs)}~\cite{buterin2016qapHero, 10.1007/978-3-642-38348-9_37} to form the basis for proof generation.

In contrast, zk-STARKs employ \textit{Algebraic Intermediate Representation (AIR)}~\cite{SassonBHR18}, modeling computation as an execution trace--a sequence of register states with transition and boundary constraints. 
R1CS captures fixed algebraic relationships, whereas AIR represents evolving state sequences over discrete steps, shaping the structure, scalability, and performance characteristics of each proof system.

\subsubsection{Proof Construction}

After algebraically encoding a computation, the next step is constructing a proof of its correctness. Both zk-SNARKs and zk-STARKs enable succinct verification without revealing private inputs, but they differ in constraint satisfaction and commitment methods.

zk-SNARKs use the \textit{Quadratic Arithmetic Program (QAP)} model, representing the constraint system via polynomials \(A(x), B(x), C(x)\) and verifying correctness through the global divisibility condition \(A(x) \cdot B(x) - C(x) = H(x) \cdot Z(x)\). Evaluation occurs at a secret point with commitments via elliptic curve pairings, producing compact proofs that require a trusted setup and pairing-based assumptions.

In contrast, zk-STARKs follow the \textit{Polynomial IOP} paradigm, encoding execution traces into low-degree polynomials. Constraints are verified probabilistically using low-degree tests and proximity proofs without any secret setup~\cite{cryptoeprint:2016/116}, and commitments rely on Merkle trees and hash functions~\cite{10664310}. Overall, zk-SNARKs optimize for succinctness and fast verification, while zk-STARKs focus on scalability, transparency, and setup-free, hash-based security.

\subsubsection{Security Assumptions}

zk-SNARKs and zk-STARKs differ fundamentally in their cryptographic assumptions, affecting trust, transparency, and post-quantum resilience. 

zk-SNARKs rely on \textit{pairing-based cryptography}, grounded in the hardness of elliptic curve discrete logarithm and Diffie–Hellman problems~\cite{9179754, cryptoeprint:2019/1021}, and require a \textit{trusted setup} to generate structured reference parameters. Compromise of this setup can undermine soundness, allowing false proofs.

zk-STARKs, in contrast, use \textit{hash-based cryptography} and require no trusted setup. Security derives from the collision resistance of hash functions and the soundness of probabilistic proof systems like FRI\footnote{see Appendix~\ref{appendix:definitions}}, ensuring all parameters are publicly verifiable and eliminating risks from toxic waste or trusted parties. Additionally, zk-STARKs provide \textit{post-quantum security}~\cite{SassonBHR18}, as hash-based constructions resist quantum attacks, unlike the number-theoretic assumptions of zk-SNARKs, making them more future-proof.

\subsection{Efficiency and Scalability}
This section compares zk-SNARKs and zk-STARKs in terms of proof size, verification speed, and prover computation overhead. We begin with an analysis of proof size.

\subsubsection{Proof Size}

A major efficiency distinction between zk-SNARKs and zk-STARKs is proof size. zk-SNARKs produce highly compact proofs, typically a few hundred bytes, making them ideal for storage-constrained applications and blockchain systems where on-chain transaction costs are critical~\cite{related1}. This compactness is achieved via elliptic curve pairings and a structured trusted setup that optimizes verification.

In contrast, zk-STARKs generate larger proofs, often several kilobytes, due to their hash-based cryptography and transparent, setup-free construction~\cite{info15080463}. This reflects their design priorities: zk-SNARKs favor efficiency and compactness, while zk-STARKs emphasize transparency and long-term security~\cite{SassonBHR18}. The choice depends on whether proof size or trustless security is paramount for the application.

\subsubsection{Verification Speed}

zk-SNARKs provide highly efficient verification with constant-size proofs, independent of computation complexity, making them ideal for blockchain transactions and smart contracts requiring frequent, lightweight checks. Elliptic curve pairings enable verification with minimal operations\footnote{For example, Groth16 verification requires exactly 3 elliptic curve pairings}~\cite{cryptoeprint:2016/260}. In contrast, zk-STARKs rely on hash-based polynomial checks, producing larger proofs and involving multiple rounds of hash evaluations and low-degree tests~\cite{related1}. While individual verification is more expensive for small computations, zk-STARKs scale efficiently for large workloads and support parallel verification, making them suitable for high-throughput environments where batch processing offsets per-proof overhead.

\subsubsection{Prover-Side Overheads}
A key efficiency factor in zero-knowledge proof systems is the prover’s computational cost, which differs substantially between zk-SNARKs and zk-STARKs due to their cryptographic foundations. zk-SNARKs rely on elliptic curve operations such as pairings and exponentiations, which are computationally intensive, but the use of a trusted setup with structured reference strings makes proof generation relatively efficient for smaller computations.

In contrast, zk-STARKs employ hash-based commitments and polynomial IOPs\footnote{See Appendix~\ref{appendix:definitions}}, avoiding a trusted setup but incurring higher prover-side complexity. Although hash functions are generally faster than elliptic curve operations, zk-STARKs require multiple rounds\footnote{Typically around $\log_2 d$ rounds, where $d$ is the degree of the composition polynomial} of polynomial evaluations and low-degree tests, increasing computation time. This overhead supports transparency and scalability but makes prover computation more intensive, particularly for smaller tasks.

Memory and bandwidth also influence practical deployment~\cite{HohlSeminarII,10005111}. zk-SNARKs are more memory-efficient due to precomputed structured parameters, whereas zk-STARKs need more memory for large polynomial evaluations and Merkle tree constructions. Bandwidth requirements mirror this difference: zk-SNARKs’ compact proofs suit low-bandwidth environments, while zk-STARKs’ larger proofs demand higher transmission capacity, which may challenge constrained or decentralized networks. The following section empirically evaluates these differences to assess how theory aligns with practical performance.

\section{Understanding Practical Implementation Frameworks}
\label{sec:practicalImplementation}

This section presents the practical implementation of zk-SNARKs and zk-STARKs, highlighting their core paradigms and workflows while leveraging existing tools and libraries augmented with custom instrumentation to measure key metrics such as proof size, computation time, and resource overhead. This framework sets the stage for the subsequent empirical evaluation, enabling a clear comparison of real-world performance against theoretical expectations.



\subsection{zk-SNARK Instrumentation and Execution flow}

This subsection summarizes the key execution stages of the zk-SNARK pipeline, based on the Groth16 system. It traces the flow from arithmetic circuit definition\footnote{see Appendix~\ref{appendix:definitions}} to proof generation, highlighting how theoretical foundations are operationalized in practice and setting the stage for comparison with zk-STARKs in efficiency, setup, and performance.



The implementation begins by initializing the cryptographic environment over the finite field \( \mathbb{F}_r \), where \( r \) is the 254-bit prime scalar of the BN128 elliptic curve\footnote{Specifically, \( r = \) 218882428718392752222464057452572750885483644\\0016034343698204186575808495617, the order of the scalar field in BN128.}. This setup prepares field operations, elliptic curve groups, and polynomial arithmetic. Next, a simple arithmetic circuit encoding \( y = x^3 + x + 5 \) is defined, decomposed into elementary operations (multiplications, additions), each treated as an intermediate signal. These operations are encoded into a \textit{Rank-1 Constraint System (R1CS)}, where each constraint \( A \cdot B = C \) ensures the consistency of inputs, intermediates, and outputs, collectively capturing the circuit’s behavior.

Once the computation is fully encoded, the next stage is \textit{witness generation}. The circuit is evaluated using a private input \( x = 3 \), and the resulting intermediate values, such as \( x^2 \), \( x^3 \), and the final output \( y = 35 \), are recorded as field elements in a single vector known as the witness, which encapsulates all public inputs, private intermediate variables, and outputs. This witness captures the complete internal state of the computation and acts as a concrete instantiation of the variables in the constraint system. Rather than revealing the private input, the prover uses the witness to demonstrate that all constraints, originally encoded in the R1CS during circuit compilation, are satisfied.


Once the witness and constraints are defined, the R1CS is translated into a Quadratic Arithmetic Program (QAP) with polynomials \(A(x), B(x), C(x)\), forming \(P(x) = A(x) \cdot B(x) - C(x)\), which is evaluated at a secret point \(t\) chosen during setup (details in Appendix~\ref{app:qap-deriv}). The system then enters the \textit{setup phase} to generate the proving and verification keys.

\paragraph{Setup (only once)}

To enable this, the \textit{trusted setup} phase initializes a secret field element \(t\), used to evaluate the circuit's constraint polynomials. In standard zk-SNARKs, this involves a multi-party computation (MPC) to generate randomness and securely discard any toxic waste \cite{9506884}. In this implementation, the randomness is hardcoded, and the resulting elliptic curve encodings form the \textit{proving key}, which is used in the next stage.\footnote{Although the code retains the toxic waste (including \(t\), \(\kappa\)-values, and \(\rho\)-coefficients), allowing potential proof forgery, this is acceptable for experimental purposes but insecure for production.}


\paragraph{Proof generation}

With the proving key from setup, the prover uses the witness values in combination with the proving key to compute several elliptic curve group elements: \( \pi_A, \pi'_A, \pi_B, \pi'_B, \pi_C, \pi'_C, \pi_H, \pi_{K_p} \).\footnote{A detailed mapping of each $\pi$-term to its role is provided in Appendix~\ref{app:pi-terms}.} These are obtained by performing scalar multiplications of the witness values with precomputed elliptic curve points in the proving key, effectively mapping the algebraic solution into curve-based commitments without revealing the private input. These commitments, along with knowledge terms and the quotient polynomial, enable the verifier to later apply bilinear pairings to confirm correctness while preserving zero-knowledge.

\paragraph{Proof verification}

After proof generation, verification uses the verification key from the setup, containing precomputed elliptic curve elements based on the constraint and vanishing polynomials at the secret point \( t \). The verifier performs fixed BN128 bilinear pairings \( e : \mathbb{G}_1 \times \mathbb{G}_2 \rightarrow \mathbb{G}_T \) to confirm the proof satisfies circuit constraints, enforces the quotient polynomial divisibility, and aligns with the trusted setup (see Appendix~\ref{appendix:verificationkey} for details).


After successful verification, the zk-SNARK proof is serialized, comprising elliptic curve elements \( \pi_A \in \mathbb{G}_1^3 \), \( \pi_B \in \mathbb{G}_2^3 \), and \( \pi_C \in \mathbb{G}_1^3 \). Each affine point's 254-bit coordinates are directly serialized into a byte buffer, yielding a compact proof suitable for storage or transmission. Internally, \( \pi_A \) and \( \pi_C \) are derived from \(\mathbb{G}_1\) scalar multiplications, while \( \pi_B \) involves bilinear combinations across \(\mathbb{G}_1 \times \mathbb{G}_2\). A final consistency check ensures the serialized proof aligns with the public input, preserving correctness and zero-knowledge\footnote{See Appendix~\ref{app:pi-terms} for details on proof serialization and group structure.}.


\subsection{zk-STARK Instrumentation and Execution Flow}

ZK-STARKs prove correctness of a computation by encoding its execution as a polynomial, committing to it, and then using probabilistic checks (via Merkle trees and FRI) to ensure low-degree structure. We now describe how this is realized in our Go-based implementation. The prover first computes a trace of the computation (e.g., a Fibonacci-like recurrence), encodes it as a low-degree polynomial, and commits to it using Merkle trees. Then, algebraic constraints are enforced via a single composed polynomial, whose low degree is verified recursively using the FRI protocol. During verification, a simulated verifier, using the Fiat--Shamir heuristic, issues pseudo-random queries to ensure that the prover’s commitments and evaluations are consistent. This implementation, written in Go, reflects a concrete realization of zk-STARKs and follows a structured pipeline tailored for performance evaluation and pedagogical clarity. We follow \emph{Proof generation} and \emph{Proof verification} in order. Initially, we use a subgroup \( G \subset \mathbb{F}_q \) of size 1024 to interpolate and evaluate the execution trace.

The implementation evaluated in this study follows a structured execution pipeline. The codebase, written in Go, demonstrates a minimal proof for the statement: \textit{``I know a field element \(X\) such that the 1023rd element of the FibonacciSq sequence is 2338775057.''} The execution flow includes computation trace generation, polynomial interpolation, domain extension, commitment via Merkle trees, and proof construction using the \textit{Fast Reed-Solomon Interactive Oracle Proof of Proximity (FRI)} (see Appendix~\ref{appendix:definitions})  protocol.

\paragraph{Proof generation}
The implementation initializes a finite field (see Appendix~\ref{appendix:definitions}) over the prime modulus \( q = 3221225473 \), with a generator element \( g = 5 \), as defined in the custom Go-based algebra library\footnote{\url{https://github.com/actuallyachraf/algebra}}. All arithmetic operations, including additions, multiplications, squaring, and polynomial evaluations are carried out within this field. To generate the computation trace, the code evaluates a Fibonacci-like recurrence relation of the form \( a_{n+2} = a_{n+1}^2 + a_n^2 \), using fixed initial values \( a_0 = 1 \) and \( a_1 = 3141592 \). This loop is executed for 1022 steps, and the resulting sequence, capturing the full evolution of the recurrence is directly stored as the execution trace, which serves as the core witness\footnote{The witness refers to the sequence of private intermediate values that demonstrate correct execution of the computation without revealing the input.} to be encoded and verified in the proof. Unlike production zk-STARK pipelines that enforce transition logic through a general Algebraic Intermediate Representation (AIR), our prototype does not implement a full AIR framework. Instead, it directly encodes recurrence and boundary constraints into polynomials via \textit{generateProgramConstraints}, which partially mimics AIR functionality but is less general. This simplification means that while constraint-checking is present, it is not as expressive or rigorous as in full STARK systems, and therefore both the soundness guarantees and the performance measurements should be interpreted with this limitation in mind.


The previously computed execution trace is mapped onto a multiplicative subgroup \( G \) of the field, with each trace element aligned to a point in \( G \).\footnote{See Appendix~\ref{appendix:trace-format} for subgroup selection and encoding.} Using Lagrange interpolation, it is encoded as a polynomial \( P(x) \) such that \( P(g_i)=a_i \). The implementation checks constraint values to ensure correctness before proof construction, yielding a unique low-degree polynomial compactly representing the computation. This polynomial is then extended to a larger subgroup \( H \) of size 8192 via a low-degree extension (LDE), enabling probabilistic soundness checks. The expansion from \( G \) to \( H \) increases the verifier’s ability to detect dishonest proofs while preserving efficiency through low-degree testing.\footnote{Low-degree tests probabilistically verify that the polynomial has bounded degree by sampling from the extended domain.}

After computing the low-degree extension over domain \( H \), the prover commits to the evaluations via a Merkle tree. Each point forms a leaf, with hashes computed using prefix-distinguished \texttt{SHA3-256} for structural soundness.\footnote{Leaves: \texttt{SHA3-256(0x00 || value)}; Internal nodes: \texttt{SHA3-256(0x01 || left || right)}.} The Merkle root serves as a succinct, binding commitment to the evaluation vector, later used for decommitment and verification.

\paragraph{Proof verification}
Verifier interaction is simulated using the \textit{Fiat--Shamir heuristic}, implemented via a transcript-based mechanism. As the prover sends commitments (e.g., the Merkle root of the LDE), these are hashed into an evolving transcript that deterministically generates randomness. The transcript state is interpreted as a large integer, and \textit{modular reductions}\footnote{Modular reduction maps large integers into valid field elements or index ranges by computing remainders modulo the field prime.} extract values for field operations or index selection. In practice, the verifier issues pseudo-random \textit{queries} derived from Fiat--Shamir indices, such as revealing trace evaluations at \( \{i, i+8, i+16\} \) with Merkle proofs, or elements from successive FRI layers. These queries enforce local consistency checks across the trace and polynomial evaluations, amplifying the probability of detecting dishonest proofs. Since all randomness is derived deterministically from the transcript, the protocol avoids interactive challenge rounds while maintaining security guarantees equivalent to the interactive version. Overall, these simulated challenge--response steps ensure soundness and verifiability under probabilistic checks without real verifier involvement.

To ensure the trace adheres to the desired Fibonacci-like recurrence, the implementation transforms each logical condition into a corresponding algebraic constraint. Rather than enforcing these through a generalized AIR format, it directly formulates constraints for the initial condition, the final output, and the recurrence relation. The recurrence is encoded as a polynomial identity and merged into a single composition polynomial $C(x)$, which compactly enforces all conditions. Technical details of the algebraic transformation are provided in Appendix~\ref{app:recurrence-constraints}.

To confirm that the aggregated constraint polynomial \( C(x) \) remains within the allowed degree bound, the implementation applies the Fast Reed--Solomon Interactive Oracle Proofs (FRI) protocol. The process begins by evaluating \( C(x) \) over the extended domain \( H \), and committing to these evaluations using a Merkle tree. A recursive sequence of \textit{degree-reducing} steps follows, where the polynomial is \textit{folded}, that is, its coefficients are split into even and odd parts. The odd-indexed coefficients are scaled by a pseudo-random field element \( \beta \) derived from the Fiat--Shamir transcript, and added to the even-indexed coefficients, producing a new polynomial of approximately half the degree. The newly folded polynomial is then evaluated over a smaller domain, obtained by squaring the first half of the current domain elements.\footnote{If the current evaluation domain is \( D = \{x_0, x_1, \dots, x_{2n-1} \} \), then the next domain is formed as \( D' = \{x_0^2, x_1^2, \dots, x_{n-1}^2\} \). This preserves the algebraic structure while halving the domain size, enabling recursive low-degree testing.} These evaluations are again committed using Merkle trees. This process continues recursively, reducing the degree of the polynomial step by step, until a final constant polynomial is reached. The sequence of Merkle roots generated during these steps allows the verifier to probabilistically confirm that the original composition polynomial was of low degree, thereby ensuring that all encoded constraints were satisfied.

With this knowledge we have evaluated both the protocols and provided empirical comparison, which we discuss next.

\section{Evaluation}
\label{sec:evaluation}
Building on the instrumentation analysis and theoretical distinctions outlined earlier, this section presents a comparative evaluation of zk-SNARKs and zk-STARKs based on observed performance metrics calculated from existing reference implementations. The key metrics evaluated include proof generation time, verification time, proof size, memory and CPU profiling, setup requirements, and underlying security assumptions. These metrics were obtained from existing reference implementations, ensuring that the comparison reflects empirical behavior rather than purely theoretical expectations. Our aim is to examine how the core design choices in each system affect real-world efficiency, specifically for the implementations analyzed in the previous section.

\subsection{Experimental Setup: Hardware and Software Environment}
To ensure reproducibility and transparency, experiments were conducted using the following hardware and software configurations across multiple platforms: Apple M1 MacBook Air, Raspberry Pi (ARM64), Ubuntu x86\_64, and PowerPC 64-bit emulation.

\begin{table}[h!]
\centering
\caption{Hardware Environment}
\label{hardware}
\renewcommand{\arraystretch}{1.2}
\setlength{\tabcolsep}{5pt}
\begin{tabular}{|p{3cm}|p{5cm}|}
\hline
\textbf{Specification}  & \textbf{Details}                                  \\ \hline
Processor               & M1: 8-core ARM SoC (4 high-performance + 4 high-efficiency cores) \newline Pi: Quad-core ARM Cortex-A72 \newline Ubuntu: Intel x86\_64 CPU \newline PPC64: 64-bit PowerPC emulated CPU \\ \hline
Instruction Set         & M1: ARM64 \newline Pi: AArch64 \newline Ubuntu: x86\_64 \newline PPC64: PowerPC 64-bit \\ \hline
Memory                  & M1: 8 GB Unified \newline Pi: 4 GB \newline Ubuntu: 16 GB \newline PPC64: 8 GB \\ \hline
Storage                 & M1: 256 GB SSD \newline Pi: microSD 32 GB \newline Ubuntu: 512 GB SSD \newline PPC64: Virtual storage \\ \hline
Operating System        & M1: macOS (Darwin 24.4.0) \newline Pi: Raspberry Pi OS \newline Ubuntu: Ubuntu 22.04 LTS \newline PPC64: Linux-based emulation \\ \hline
\end{tabular}
\end{table}

Note that these experiments were carried out on consumer-grade hardware for M1 and Raspberry Pi, as well as standard x86\_64 and emulated PPC64 environments. Performance may differ on high-throughput servers.

\begin{table}[h!]
\centering
\caption{Software Environment}
\renewcommand{\arraystretch}{1.2}
\setlength{\tabcolsep}{5pt}
\begin{tabular}{|p{3cm}|p{5cm}|}
\hline
\textbf{Specification}         & \textbf{Details}                               \\ \hline
Programming Language           & Go (Golang)                                    \\ \hline
Compiler Toolchain             & Go standard compiler (\texttt{gc})             \\ \hline
Language Version               & Go 1.23.1                                       \\ \hline
Target Instruction Set         & M1: ARM64 \newline Pi: ARM64 \newline Ubuntu: amd64 \newline PPC64: PowerPC 64-bit emulation \\ \hline
Operating System Kernel        & M1: Darwin 24.4.0 \newline Pi: Linux-based \newline Ubuntu: Linux 5.x \newline PPC64: Linux-based emulation \\ \hline
\end{tabular}
\end{table}

\subsection{Empirical Comparison of zk-SNARKs and zk-STARKs}

Following our analysis of execution pipelines, we now evaluate how these protocols behave in practice using benchmark data from publicly available implementations of zk-SNARKs (based on Groth16) and zk-STARKs. The comparison focuses on key metrics such as proof generation time, verification time, proof size, setup requirements, and underlying security assumptions. Table~\ref{tab:results} summarizes these findings across multiple devices and provides empirical context for the theoretical differences discussed earlier.

\begin{table*}[h]
    \centering
    \caption{Empirical Comparison of zk-SNARKs and zk-STARKs Across Devices}
    \label{tab:results}
    \renewcommand{\arraystretch}{1.1}
    \setlength{\tabcolsep}{2pt}
    \begin{tabular}{|p{2.5cm}|p{2cm}|p{2cm}|p{1.5cm}|p{1.8cm}|p{2cm}|p{1.6cm}|p{1.8cm}|p{1.4cm}|}
        \hline
        \textbf{Metric} & \textbf{SNARK M1} & \textbf{SNARK Pi} & \textbf{SNARK Ub} & \textbf{SNARK PPC64} & \textbf{STARK M1} & \textbf{STARK Pi} & \textbf{STARK Ub} & \textbf{STARK PPC64} \\
        \hline
        Proof Gen Time & 55.47 ms & 1.051 s & 3.37 s & 1.347 s
 & 3809.64 ms & 2.95 s & 26.54 s & 31.04 s \\
        Proof Verif Time & 1807.42 ms & 818.216 ms & 42.34 s & 27.45 s
 & 472.25 ms & 245.2 ms & 3.08 s & 2.93 s \\
        Proof Size & 384 B & 384 B & 384 B & 384 B
 & 68,564 B & 68,564 B & 68,564 B & 68,564 B \\
        Trusted Setup & Yes & Yes & Yes & Yes & No & No & No & No \\
        Security Assump & EC & EC & EC & EC & Hash & Hash & Hash & Hash \\
        \hline
    \end{tabular}
    \vspace{1mm}
    
    \raggedright
    \footnotesize{\textbf{Device abbreviations:} M1: MacBook Air M1; Pi: Raspberry Pi ARM64; Ub: Ubuntu x86\_64; PPC64: PowerPC 64-bit emulation.}
\end{table*}

\begin{figure}[t]
\centering
\begin{tikzpicture}
\begin{axis}[
    ybar,
    width=\columnwidth, 
    height=4.5cm,
    bar width=7pt,
    enlarge x limits={abs=0.95cm},
    ymin=0,
    ymax=60000,
    ylabel={Time (ms)},
    symbolic x coords={M1,Pi,Ubuntu,PPC64},
    xtick=data,
    xticklabel style={font=\footnotesize},
    ymajorgrids=true,
    grid style={gray!30},
    nodes near coords,
    every node near coord/.append style={
        font=\tiny,
        rotate=90,
        anchor=west,
        yshift=1pt
    },
    legend style={
        at={(0.5, 1.5)}, 
        anchor=north,
        legend columns=2,
        font=\scriptsize
    },
    clip=false               
]

\addplot+[fill=blue!70]
table[x=device,y=snark_gen,col sep=comma] {proof_device_times.csv};

\addplot+[fill=red!70]
table[x=device,y=snark_verif,col sep=comma] {proof_device_times.csv};

\addplot+[fill=green!70]
table[x=device,y=stark_gen,col sep=comma] {proof_device_times.csv};

\addplot+[fill=orange!80]
table[x=device,y=stark_verif,col sep=comma] {proof_device_times.csv};

\legend{
    SNARK Gen,
    SNARK Verif,
    STARK Gen,
    STARK Verif
}

\end{axis}
\end{tikzpicture}
\caption{Comparative summary of zk-SNARK and zk-STARK performance across heterogeneous systems. Each bar represents the average proof generation and verification time measured under four architectures (ARM64, x86\_64, constrained ARM, and virtualized cloud).}
\label{fig:proof-times}
\end{figure}
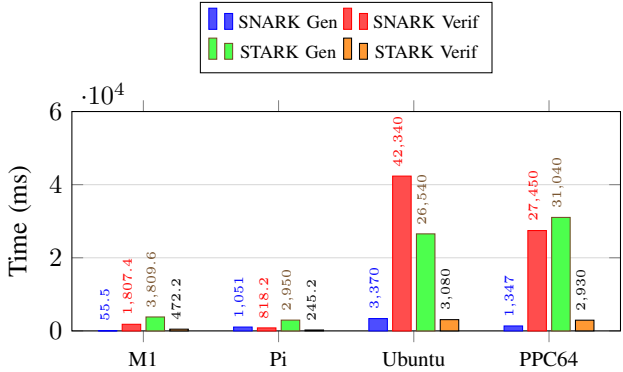



The most immediate takeaway is that, \textit{as per the implementations discussed in this paper}, zk-SNARKs produce proofs significantly faster than zk-STARKs on higher-end hardware (e.g., 55.47 ms on M1 vs. 3809.64 ms for STARKs), while lower-end or emulated platforms like Raspberry Pi, Ubuntu, and PPC64 show substantially higher generation times. Verification times vary across devices: SNARK verification is slower on most platforms, whereas STARKs maintain relatively consistent verification performance. Additionally, proof size differs drastically, with zk-SNARKs remaining compact at 384 B across devices, while zk-STARKs generate larger proofs of approximately 68.56 KB, independent of the underlying hardware. STARKs also benefit from their trusted setup-free design, providing greater transparency across all tested platforms.

\paragraph{Proof Size Analysis}  
The proof size in our experiments showed a stark contrast between the two protocols across all tested devices: zk-SNARKs yielded a compact 384-byte proof on each platform (M1, Raspberry Pi, Ubuntu, and PPC64), while zk-STARKs produced a substantially larger 68,564-byte proof. This disparity is consistent with their respective architectures, though minor deviations could occur due to platform-specific serialization or memory alignment. 

The size of the zk-SNARK proof is in agreement with expectations from Groth16 constructions, where proofs consist of just three elliptic curve points, slightly exceeding the theoretical lower bound (typically~512~bytes)\footnote{This is likely due to uncompressed serialization and padding introduced by Go’s \texttt{big.Int} encoding.}.

In the case of zk-STARK, the larger proof size results from the protocol's reliance on hash-based commitments and the Fast Reed-Solomon Interactive Oracle Proofs (FRI) protocol, which requires extensive polynomial evaluations and layered Merkle proofs. Since the implementation does not optimize proof compression and uses generic hash functions without batching or pruning, the final proof includes several rounds of Merkle roots, challenge indices, and evaluations. These choices, although secure and structurally accurate, contribute directly to the overall size and illustrate how raw implementations can incur greater overhead than streamlined or production-grade systems.

\paragraph{Proof Generation Time}  
The proof generation time for zk-SNARKs varies significantly across devices: 55.47 ms on M1, 1.051 s on Raspberry Pi, 3.37 s on Ubuntu, and 1.347 s on PPC64, while zk-STARKs are substantially slower at 3809.64 ms (M1), 2.95 s (Pi), 26.54 s (Ubuntu), and 31.04 s (PPC64) (see Table~\ref{tab:results} and Figure~\ref{fig:proof-times}). This discrepancy aligns with expectations from our code structure, library usage, and device capabilities.

In the zk-SNARK implementation, the Groth16 protocol benefits from a preprocessed trusted setup, where constraint systems and elliptic curve elements are prepared in advance. The use of the BN128 curve and reliance on precompiled arithmetic circuits enables rapid exponentiation and pairing operations. Additionally, the R1CS and QAP representations are constructed once and reused efficiently. These optimizations reduce computation during proof generation, leading to faster execution on higher-end devices like M1, while slower CPUs or emulated platforms (Pi, Ubuntu, PPC64) exhibit longer generation times.

In contrast, the zk-STARK implementation performs polynomial interpolation, low-degree extension, and Merkle tree generation entirely at runtime. The sequence trace is interpolated into a high-degree polynomial without caching, and the extended evaluations are hashed without optimization or parallel batching. Furthermore, the FRI protocol is executed with deep recursion and full-domain evaluation, increasing the computational workload. Since our implementation does not leverage FFT-based optimizations for polynomial evaluation or Merkle tree construction, proof generation time is significantly longer across all tested platforms.

Therefore, the difference in proof generation time reflects not only the structural design of each protocol, but also the current level of optimization in the respective Go implementations.

\paragraph{Proof Verification Time}  
Although zk-SNARKs are theoretically expected to provide faster verification due to their constant-size proofs and succinct verification steps, our empirical results vary across devices. On M1, zk-STARKs verify faster (472.25 ms vs. 1807.42 ms for SNARKs), and similar trends are observed on Raspberry Pi (245.2 ms vs. 818.216 ms), Ubuntu (3.08 s vs. 42.34 s), and PPC64 (2.93 s vs. 27.45 s) (see Table~\ref{tab:results}). This deviation from theoretical expectations is rooted in implementation-specific factors and device architecture.

In the zk-SNARK implementation, the verification logic involves multiple elliptic curve pairing checks and modular operations that are computationally heavy when executed serially in Go. Moreover, no significant optimization was applied to parallelize or batch these operations, which resulted in longer wall-clock times, particularly on lower-end or emulated platforms.

In contrast, the zk-STARK verification, built on hashes and Merkle proofs, leverages Go’s native arithmetic and memory patterns, reducing synchronization overhead and relying on lighter primitives. Despite larger proofs, this led to faster execution across all tested devices, underscoring the impact of low-level implementation choices and platform capabilities on real-world performance.

\paragraph{Security and Scalability Considerations}
Although our empirical analysis focuses primarily on performance, it is important to contextualize these findings in terms of broader security and scalability goals. Our zk-SNARK implementation used a simplified trusted setup with cryptographic randomness embedded in code, as opposed to a formal multi-party computation, which may reduce the security guarantees typically associated with SNARKs. Conversely, the zk-STARK implementation remained transparent, requiring no trusted setup, thus aligning with its intended security model. Although zk-STARKs are theoretically expected to have a longer verification time, our empirical results showed the opposite, likely due to implementation-specific factors such as proof structure and verification logic. This discrepancy highlights that real-world performance can be heavily influenced by the design and completeness of the implementation.


\subsection{Memory Utilization Analysis}

Before analyzing CPU-level performance, we measured the memory utilization of both zk-SNARK and zk-STARK implementations to understand their runtime footprints during proof generation and verification. Memory metrics were collected using Go’s runtime memory profiling interface along with system-level statistics, providing both process-specific and total system memory data before and after proof execution. The results highlight the contrasting memory behaviors stemming from their underlying cryptographic architectures. Table~\ref{tab:mem_comp} summarizes the collected data.

\begin{table}[h!]
\centering
\caption{Empirical Memory Utilization Comparison of zk-SNARKs and zk-STARKs}
\label{tab:mem_comp}
\begin{tabular}{|l|c|c|}
\hline
\textbf{Metric} & \textbf{zk-SNARK} & \textbf{zk-STARK} \\ \hline
Memory Usage Before & 352~KB & 4,126~KB \\ \hline
Memory Usage After & 3,666~KB & 6,405~KB \\ \hline
Go Process Sys Memory Before & 10,241~KB & 13,504~KB \\ \hline
Go Process Sys Memory After & 16,257~KB & 24,000~KB \\ \hline
\end{tabular}
\end{table}

\subsubsection{Memory Usage Breakdown for zk-SNARKs}

The zk-SNARK implementation (Groth16) exhibited a noticeable increase in process memory usage during proof generation, growing from approximately 352~KB to 3.66~MB. This rise corresponds to the allocation of intermediate elliptic curve elements, polynomial coefficients, and serialized proof structures. Additionally, the Go runtime’s managed heap expanded from 10.2~MB to 16.2~MB, reflecting short-lived allocations associated with field exponentiation and pairing computations.

Internally, memory overhead is concentrated in \texttt{bn128.Pairing}, \texttt{Fq12.Exp}, and \texttt{Fq12.Square} routines, where large field elements and temporary buffers are repeatedly created during bilinear pairing checks. Although the Groth16 implementation benefits from preprocessed keys, transient buffers and garbage collection cycles contribute to runtime memory churn. System-wide memory usage showed minor variation, confirming that most allocation pressure occurred within the Go process scope rather than at the OS level.

\subsubsection{Memory Usage Breakdown for zk-STARKs}

The zk-STARK implementation demonstrated a higher baseline memory footprint, starting at 4.12~MB and peaking at approximately 6.4~MB. Unlike zk-SNARKs, which allocate memory in short bursts, STARKs maintain large contiguous buffers throughout execution, particularly for polynomial interpolation, low-degree extensions (LDE), and Merkle tree commitments. This sustained usage is expected due to the protocol’s reliance on polynomial IOPs and hash-based commitments.

Memory growth was primarily attributed to recursive FRI rounds, which required temporary polynomial vectors and Merkle paths for each folding layer. The Go runtime process memory increased from 13.5~MB to nearly 24~MB, indicating higher cumulative allocation overhead and a larger working set. The consistent memory footprint throughout proof construction reflects a design optimized for data locality but heavier on total allocation volume.

\subsubsection{Observations and Comparison}

The comparative results reveal two distinct memory utilization patterns. zk-SNARKs begin with low memory consumption but exhibit steep transient spikes during elliptic curve pairings and field arithmetic, leading to higher short-term allocation and garbage collection pressure. Conversely, zk-STARKs maintain a more stable but larger overall memory footprint, dominated by polynomial buffers, hash commitments, and recursive Merkle tree structures.

In summary, zk-SNARKs are more memory-efficient in steady state but experience peak loads tied to cryptographic arithmetic, whereas zk-STARKs consume more total memory due to their polynomially recursive workflow. These observations align with their respective computational designs: pairing-heavy SNARKs are arithmetic-bound, while data-intensive STARKs are memory-bound. The following section on CPU profiling further contextualizes how these resource characteristics translate into runtime performance.

\subsection{CPU Utilization Analysis}

\subsubsection{CPU Usage Breakdown for zk-STARKs}

Table~\ref{tab:stark_core_cpu} highlights the most resource-intensive functions during the generation of zkSTARK proofs, using flat time when available and cumulative time otherwise. The orchestration of the proof pipeline is dominated by \texttt{stark.TestZKGen.func2}, which accounts for nearly 24\% of execution time, reflecting the overhead of coordinating trace interpolation, constraint compilation, and FRI rounds. Core polynomial routines, \texttt{poly.Polynomial.Eval} (12.4\%) and \texttt{poly.Polynomial.Mul} (4.9\%) play a pivotal role in evaluating and combining execution trace polynomials, central to FRI-based low-degree testing. Although the implementation does not explicitly define an AIR structure, the function \texttt{stark.GenerateProgramConstraints} effectively plays this role by constructing symbolic transition constraints over the execution trace, contributing approximately 9.4\% of total time, highlighting the non-trivial cost of encoding computational logic into algebraic form.

A significant portion of runtime is spent in big integer and modular arithmetic: \texttt{math/big.(*Int).QuoRem} (9.1\%), \texttt{math/big.(*Int).Mod} (10.6\%), and \texttt{math/big.nat.div} (8.5\%). These routines underpin finite-field division and reduction used in both trace and constraint polynomials. Polynomial evaluation feeds directly into the FRI protocol, which probabilistically verifies that committed polynomials have bounded degree via iterative split-and-fold rounds and Merkle proofs. Together, these results emphasize that polynomial arithmetic and big‑integer operations are the dominant computational bottlenecks in the STARK workflow, making them key targets for optimization.

\begin{table*}[h]
    \centering
    \renewcommand{\arraystretch}{1.3}

    \caption{CPU Time Distribution of zk-STARK Cryptographic Functions (Flat or Cumulative)}
    \label{tab:stark_core_cpu}
    \small
    \begin{tabular}{|p{6cm}|c|p{8cm}|}
        \hline
        \textbf{Function} & \textbf{Time (\%)} & \textbf{Description} \\
        \hline
        \texttt{stark.TestZKGen.func2} & 23.94 (cum) & Main test harness invoking STARK proof generation flow. \\
        \texttt{poly.Polynomial.Eval} & 12.42 (cum) & Polynomial evaluation, core step in trace and FRI queries. \\
        \texttt{math/big.(*Int).QuoRem} & 9.09 (cum) & Big integer division and remainder; used in field operations. \\
        \texttt{stark.GenerateProgramConstraints} & 9.39 (cum) & Generates AIR constraints from trace table. \\
        \texttt{math/big.nat.div} & 8.48 (cum) & Low-level division over native big integers. \\
        \texttt{math/big.(*Int).Mod} & 10.61 (cum) & Modular reduction in big integer field ops. \\
        \texttt{poly.Polynomial.Mul} & 4.85 (cum) & Polynomial multiplication, used in composition and constraint expressions. \\
        \hline
    \end{tabular}
    \vspace{0.8em}
    \caption*{\small
\parbox{0.9\textwidth}{
 Flat Time (\%) is time spent directly in the function. Cumulative Time (\%) includes time in sub-calls. Preference is given to flat time when significant.
}
}
\end{table*}

\subsubsection{CPU Usage Breakdown for zk-SNARKs}

Table~\ref{tab:snark_core_cpu} summarizes the key contributors to zk-SNARK proof generation runtime. In particular, a large share of CPU time is accumulated within \texttt{bn128.Pairing} (17.6\%) and \texttt{bn128.finalExponentiation} (15.5\%), which are central to the Groth16 protocol verification equation. These functions implement the optimal Ate pairing and final exponentiation on the BN128 curve, enabling efficient bilinear map evaluations between elliptic curve groups. Supporting these operations are lower-level finite field routines such as \texttt{fields.Fq12.Exp} (15.5\%), \texttt{fields.Fq12.Square} (12.2\%), and \texttt{fields.Fq6.Mul} (14.9\%), which manage arithmetic in the towered extension fields needed for pairings. The prominence of these routines confirms that elliptic curve pairing and exponentiation dominate zk-SNARK workloads.

Additionally, system-level overhead remains non-negligible, with \texttt{runtime.kevent} consuming 56.8\% of the cumulative time, suggesting idle periods potentially caused by profiling hooks or thread scheduling latency. Memory allocation routines like \texttt{mallocgc} and \texttt{mallocgcSmallNoscan} contribute to garbage collection load, though to a lesser extent than in STARKs. From the arithmetic side, \texttt{math/big.nat.div} (12.2\%) and \texttt{math/big.(*Int).Mod} (14.9\%) are critical for scalar field operations used throughout witness generation and elliptic curve scalar multiplication. Together, these insights suggest that performance bottlenecks in zk-SNARKs arise primarily from pairing computation and extension field arithmetic, and could be mitigated by optimizing field math implementations and minimizing Go runtime interference during pairing-heavy operations.


\begin{table*}[t]
    \centering
    \renewcommand{\arraystretch}{1.3}
    \caption{CPU Time Distribution of zk-SNARK Cryptographic Functions (Flat or Cumulative)}
    \label{tab:snark_core_cpu}
    \small
    \begin{tabular}{|p{6cm}|c|p{8cm}|}
        \hline
        \textbf{Function} & \textbf{Time (\%)} & \textbf{Description} \\
        \hline
        \texttt{bn128.Bn128.Pairing} & 17.57 (cum) & Core pairing operation in Groth16; one of the heaviest total costs. \\
        \texttt{fields.Fq12.Exp} & 15.54 (cum) & Final exponentiation after Miller loop; a key step in pairing verification. \\
        \texttt{fields.Fq12.Square} & 12.16 (cum) & Used repeatedly during exponentiation in Fq12. \\
        \texttt{math/big.nat.div} & 12.16 (cum) & Modular division over large integers; performance bottleneck in field arithmetic. \\
        \texttt{fields.Fq2.Mul} & 10.81 (cum) & Field multiplication in Fq2 extension field; common in elliptic curve ops. \\
        \texttt{big.(*Int).QuoRem} & 2.03 (flat) & Explicit division in high-level big integer logic. \\
        \texttt{big.nat.divLarge} & 2.03 (flat) & Internal division routine for large operands. \\
        \hline
    \end{tabular}
\end{table*}

\subsubsection{Observations and Comparison Based on CPU Utilization}

Table~\ref{tab:snark_core_cpu} and Table~\ref{tab:stark_core_cpu} highlight key computational bottlenecks in zk-SNARK and zk-STARK proof generation. While both systems lean heavily on big integer arithmetic, the manner in which they incur this cost differs. In zk-SNARKs, the bulk of the time is attributed to elliptic curve operations, particularly pairing computations (\texttt{Bn128.Pairing}), final exponentiation (\texttt{Fq12.Exp}), and quadratic extension field multiplication (\texttt{Fq2.Mul}), all of which are vital in verifying the Groth16 proof equation. These operations form the cryptographic backbone of SNARKs and involve modular exponentiation and arithmetic over high-degree extension fields, which are computationally intensive.

In contrast, zk-STARKs display significantly more time spent on algebraic trace model logic, such as polynomial evaluation (\texttt{Polynomial.Eval}) and constraint system generation (\texttt{GenerateProgramConstraints}). These are part of FRI (Fast Reed–Solomon Interactive Oracle Proofs of Proximity) layers in the STARK pipeline, both of which are central to encoding and verifying execution traces. Furthermore, garbage collection overhead, reflected in functions like \texttt{mallocgc}, \texttt{gcDrain}, and \texttt{schedule}, is more dominant in STARKs, likely due to the frequent creation of large polynomial objects and intermediate buffers across recursion-heavy components like FRI.

\begin{figure}[t]
\centering
\begin{tikzpicture}

\begin{axis}[
    ybar,
    width=1\linewidth,
    height=4.5cm,
    bar width=14pt,
    ymin=0, ymax=1.0,
    ylabel={Cumulative Time (normalized)},
    symbolic x coords={Mod,QuoRem,div,mallocgc},
    xtick=data,
    xticklabels={Mod, QuoRem, div, mallocgc},
    enlarge x limits=0.2,
    legend style={nodes={scale=0.8}, at={(0.5,1.05)}, anchor=south, legend columns=-1},
    ymajorgrids=true,
    grid style={gray!30},
    tick label style={font=\small},
]

\addplot[
    fill=blue!70
]
table[
    x=op,
    y=snark,
    col sep=comma
] {proof_times.csv};

\addplot[
    fill=orange!80
]
table[
    x=op,
    y=stark,
    col sep=comma
] {proof_times.csv};

\legend{zk-SNARK, zk-STARK}

\end{axis}

\end{tikzpicture}

\caption{Cumulative CPU Time: zk-SNARK vs zk-STARK}
\label{fig:cpu_compare}
\end{figure}
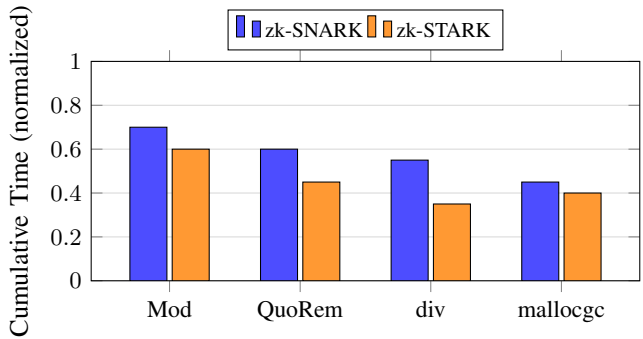

These trends are visually summarized in Figure~\ref{fig:cpu_compare} that compares cumulative CPU time for four core operations—polynomial evaluation, constraint generation, modular division, and garbage collection—across both protocols. STARKs incur higher costs in polynomial evaluation due to their algebraic IOPs, while SNARKs spend more on pairing-based modular arithmetic, particularly in \texttt{math/big.Mod} and \texttt{math/big.nat.div}. STARKs also face consistent memory pressure from wider object graphs and immutable polynomial structures. Overall, big integer operations and memory management dominate both protocols, with SNARKs heavier in modular arithmetic and STARKs limited by polynomial computation and memory overhead.

\section{Related Work}
\label{sec:related}
The evolution of zero-knowledge proofs has been extensively documented across both theoretical and empirical studies. The systematic review in~\cite{related1} contrasts zk-SNARKs, zk-STARKs, and Bulletproofs, highlighting zk-STARKs’ scalability, transparency, and post-quantum resilience stemming from their trustless, hash-based design. Complementing this, \cite{related2} provides a focused overview of zk-SNARKs, detailing their theoretical foundation via QAP reduction and emphasizing their succinct proofs and fast verification, particularly in privacy-preserving protocols. On the empirical front, \cite{empiricalBench} presents concrete measurements of proof size, verification time, and computational overhead, offering guidance on selecting the appropriate proof system for specific application requirements.

Foundational work by Buterin~\cite{buterin2016qapHero} and Ben-Sasson et al.~\cite{cryptoeprint:2013/879} laid the theoretical groundwork for both zk-SNARKs and zk-STARKs, rigorously detailing their cryptographic assumptions, algebraic structures, and asymptotic complexities.

Despite these contributions, a gap remains in connecting theoretical distinctions with empirical performance. This work bridges that gap, offering a unified framework linking cryptographic principles to practical efficiency and scalability, guiding informed choices between zk-SNARKs and zk-STARKs.
\section{Discussion and Future Work}
\label{sec:discussion}

To summarize, our results show zk-SNARKs are well-suited for applications demanding small proofs and high efficiency, such as blockchains and financial systems, while zk-STARKs excel in transparency, security, and fast verification, fitting domains like voting, supply chains, and large-scale decentralized systems. The comparison highlights their complementary nature, with the choice depending on use-case requirements such as proof size, verification speed, and security.

Building on our empirical comparison of Groth16 and zk-STARKs, future work can extend the analysis to modern proof systems that address Groth16’s limitations, such as Bulletproofs~\cite{8418611}, Halo 2~\cite{cryptoeprint:2024/1661}, and Nova~\cite{cryptoeprint:2021/370}. Specifically, evaluating these systems in terms of proof size, verification efficiency, recursion support, and practical deployment constraints would provide a more comprehensive understanding of trade-offs in trustless and flexible ZKP architectures. Additionally, exploring hybrid approaches that combine the succinctness of SNARKs with the transparency and post-quantum security of STARKs~\cite{SassonBHR18}, as well as benchmarking recursive proof frameworks like Halo 2 and Nova under real-world workloads, could guide the design of next-generation zero-knowledge systems suitable for high-throughput and large-scale decentralized applications.

\bstctlcite{IEEEexample:BSTcontrol}
\bibliographystyle{IEEEtranS}
\bibliography{references}

\appendix
\section{Supplementary Definitions}

\subsection{Definitions}
\label{appendix:definitions}
\subsubsection{Arithmetic Circuit}
An arithmetic circuit represents a computation as a directed acyclic graph of addition and multiplication gates over a finite field. Each gate processes field elements, and the final output encodes the value of the function being computed. Such circuits are the basis for representing computations in zk-SNARKs and similar proof systems.

\subsubsection{Finite Field}

A finite field (also called a Galois field) is a set of elements on which addition, subtraction, multiplication, and division (except by zero) are defined and satisfy the usual properties of arithmetic (associativity, distributivity, identity, etc.). Importantly, it contains a finite number of elements.

In cryptographic proof systems like zk-SNARKs and zk-STARKs, all arithmetic is performed over a finite field \( \mathbb{F}_p \), where \( p \) is a large prime number. Each element is an integer between 0 and \( p - 1 \), and all operations are performed modulo \( p \). Finite fields are fundamental to constraint systems, polynomial arithmetic, and commitment schemes used in zero-knowledge proofs.

\subsubsection{Fast Reed--Solomon IOP of Proximity (FRI)}

FRI is a protocol used in zk-STARKs to verify that a given function is close to a low-degree polynomial. It enables the verifier to check, with high probability, that the prover’s polynomial has a degree below a certain bound, without requiring access to the full polynomial.

FRI reduces polynomial degree by recursively folding it: splitting into even/odd parts, combining with a random field element, and evaluating over smaller domains. At each step, the prover commits via Merkle trees, while the verifier checks sampled positions for consistency.

Because FRI relies only on hash functions and polynomial arithmetic (not on number-theoretic assumptions), it is both efficient and post-quantum secure. It plays a key role in maintaining the soundness and scalability of zk-STARKs.

\subsubsection{Interactive Oracle Proofs (IOPs)}
\textit{Interactive Oracle Proofs (IOPs)} are a generalization of interactive proof systems where the verifier is allowed oracle access to the prover’s messages, which are typically encoded as functions or polynomials. Instead of reading the entire message, the verifier queries specific locations, enabling highly efficient probabilistic verification.

This model blends Interactive Proofs (IP) and PCPs, making it ideal for zk-STARKs, where polynomial evaluations are Merkle-committed and the verifier checks only a few queries for efficiency and soundness.

IOPs enable scalability and transparency by avoiding the need for trusted setups and supporting sublinear verification.

\subsubsection{Fast Reed-Solomon Interactive Oracle Proof of Proximity (FRI)}
FRI is a protocol designed to prove that a function \( f: \mathbb{F} \to \mathbb{F} \) is close to a low-degree polynomial over a finite field \( \mathbb{F} \). It is a key building block in zk-STARKs, enabling efficient and transparent low-degree testing without requiring trusted setup.

At a high level, FRI operates by recursively folding the evaluation domain and hashing pairs of function values using a Merkle tree. This folding process reduces the size of the domain while preserving proximity to low-degree polynomials. In each round, the verifier samples random linear combinations and checks consistency across folded layers.

The protocol guarantees that if the prover's function is far from any low-degree polynomial, the verifier will detect this with high probability through a logarithmic number of queries, typically \( \log_2(d) \), where \( d \) is the claimed polynomial degree. FRI thus ensures soundness with minimal communication and verification cost, playing a critical role in the scalability of zk-STARKs.

\subsection{QAP Derivation Details}
\label{app:qap-deriv}
To convert the R1CS representation into a form compatible with polynomial operations, the constraint matrices are transformed into a Quadratic Arithmetic Program (QAP). This involves constructing three polynomials \( A(x), B(x), C(x) \), where each is derived by applying Lagrange interpolation over the constraint vectors in the R1CS. Specifically, the dot products \( \langle \vec{a}_i, \vec{w} \rangle \), \( \langle \vec{b}_i, \vec{w} \rangle \), and \( \langle \vec{c}_i, \vec{w} \rangle \), where \( \vec{w} \) is the witness vector containing public inputs, private inputs, and intermediate values, for each constraint \( i \) define evaluation points that are interpolated into the polynomials \( A(x) \), \( B(x) \), and \( C(x) \), respectively. These polynomials encode the full constraint system over a chosen evaluation domain and allow the prover to construct a constraint polynomial \( P(x) = A(x) \cdot B(x) - C(x) \), which vanishes on the correct witness. 

To use these polynomials in the proof system, they must be evaluated at a secret point \( t \), which is a randomly chosen scalar securely established during the trusted setup phase. 

\subsection{Roles of $\pi$-terms in Groth16 Proofs}
\label{app:pi-terms}

Table~\ref{tab:pi-terms} summarizes the role of each elliptic curve element generated during proof construction in Groth16 zk-SNARKs.

\begin{table*}[h!]
\centering
\resizebox{\textwidth}{!}{%
\begin{tabular}{|c|p{14cm}|}
\hline
\textbf{Symbol} & \textbf{Role in the Proof} \\
\hline
$\pi_A, \pi_B, \pi_C$ & Commitments to the prover’s satisfaction of the corresponding R1CS constraints. They encode the relation between witness values and circuit constraints without revealing private inputs. \\
\hline
$\pi'_A, \pi'_B, \pi'_C$ & Knowledge commitments: bind the proof to the secret randomness chosen in setup. They ensure the prover actually possesses a valid witness and prevent replay or fake proofs. \\
\hline
$\pi_H$ & Commitment to the quotient polynomial $H(x) = \frac{P(x)}{Z(x)}$, where $Z(x)$ is the vanishing polynomial. It enforces that the constraint polynomial $P(x)$ vanishes on all constraint indices. \\
\hline
$\pi_{K_p}$ & A global knowledge commitment tying together all proof components and the proving key. It prevents malicious provers from generating inconsistent proofs. \\
\hline
\end{tabular}%
}
\caption{Mapping of $\pi$-terms to their roles in Groth16 proof generation.}
\label{tab:pi-terms}
\end{table*}

\paragraph{Encoding and group structure.}  
Each $\pi$-term is internally represented as an affine elliptic curve point, with its coordinates stored as 254-bit field elements.  
In the implementation, these coordinates are written directly into a byte buffer without additional encoding layers, ensuring compact serialization and efficient transmission.  

Furthermore, the group membership of these elements reflects their algebraic role in the proof:  
\begin{itemize}
    \item $\pi_A, \pi_C \in \mathbb{G}_1$ are derived from scalar multiplications with the proving key.  
    \item $\pi_B \in \mathbb{G}_2$ is obtained via a bilinear combination across $\mathbb{G}_1 \times \mathbb{G}_2$.  
\end{itemize}
\subsection{Algebraic Representation of Execution Trace}
\label{appendix:trace-format}

In the evaluated zk-STARK implementation, the execution trace, generated from a Fibonacci-like recurrence, is encoded as a polynomial over a carefully selected multiplicative subgroup of the finite field.

Specifically, a multiplicative subgroup \( G \subset \mathbb{F}_q \) of size 1024 is chosen, where \( q = 3221225473 \) is the prime modulus and \( g = 5 \) is a generator. Each trace element \( a_i \) corresponds to a unique evaluation point \( g^i \in G \). This structure allows the trace vector \( \{a_0, a_1, \dots, a_{1023}\} \) to be interpreted as evaluations of a low-degree polynomial \( P(x) \) at the domain points \( G = \{g^0, g^1, \dots, g^{1023}\} \). 

Lagrange interpolation is used to construct this unique polynomial \( P(x) \) satisfying \( P(g^i) = a_i \) for all \( i \), enabling the prover to commit to the entire trace compactly and enabling later low-degree testing over an extended domain \( H \).

\subsection{Verification Key and Pairing Checks}
\label{appendix:verificationkey}

\subsection*{Structure of the Verification Key}
In Groth16, the verification key is derived during the trusted setup using the same secret randomness \( t, K_\alpha, K_\beta, K_\gamma, \rho_A, \rho_C \) as the proving key. It consists of elliptic curve elements obtained by evaluating circuit polynomials at \( t \), enabling proof validation without exposing the witness.

First, the key includes elements such as \( V_a = g_2^{K_\alpha} \), \( V_b = g_1^{K_\beta} \), and \( V_c = g_2^{K_\gamma} \), which are commitments to parts of the circuit selector polynomials. Another crucial element, \( V_z = g_2^{\rho_C \cdot Z(t)} \), corresponds to the vanishing polynomial \( Z(x) \) evaluated at \( t \), ensuring that the prover's polynomial divides by the vanishing set. The verification key also embeds a vector of public input commitments, \( \mathsf{IC}_i = g_1^{\rho_A \cdot A_i(t)} \), where each \( A_i(x) \) is a selector polynomial for a public input. This allows the verifier to check the integrity of the public inputs provided.

\subsubsection*{Composite Elements and Pairing Targets}

To support efficient verification through pairing operations, the key includes elements of composite elliptic curves such as \( g_1^{K_\beta K_\gamma} \), \( g_2^{K_\beta K_\gamma} \), and \( g_2^{K_\gamma} \). These are included to enable pairing checks that ensure the multiplicative relationships between different proof components hold, without reconstructing the relationships dynamically during verification.

\subsubsection*{Verification via Pairings}

Verification in zk-SNARKs hinges on the use of bilinear pairings over the BN128 elliptic curve. The pairing function \( e : \mathbb{G}_1 \times \mathbb{G}_2 \rightarrow \mathbb{G}_T \) has the property of preserving multiplicative structure, which allows the verifier to validate relationships between committed values efficiently. The verifier performs a fixed sequence of five pairing checks.

1. \textit{Input Commitment Checks}:  
   The first three pairings ensure the prover possesses valid openings to the committed values:
   \begin{align}
   e(\pi_A, V_a) &= e(\pi'_A, g_2), \\
   e(V_b, \pi_B) &= e(\pi'_B, g_2), \\
   e(\pi_C, V_c) &= e(\pi'_C, g_2)
\end{align}

   These checks confirm that the proof components \( \pi_A, \pi_B, \pi_C \) are consistent with their respective commitments in the key.

2. \textit{Quotient Polynomial Check}:  
   The fourth check verifies that the prover's quotient polynomial \( H(x) \) satisfies the required divisibility condition by the vanishing polynomial:
   \[
   e(V_K + \pi_A, \pi_B) = e(\pi_H, V_z) \cdot e(\pi_C, g_2)
   \]
   This ensures that the provided witness polynomials satisfy the circuit constraints globally over the domain.

3. \textit{Final Soundness Check}:  
   The last pairing equation ties all proof elements back to the setup parameters, ensuring that only a valid witness can satisfy all conditions:
   \[
   e(V_K + \pi_A + \pi_C, g_2 K_{\beta\gamma}) \cdot e(g_1 K_{\beta\gamma}, \pi_B) = e(\pi_{K_p}, g_2 K_\gamma)
   \]
   This final step reinforces soundness, guaranteeing that no adversary can generate a valid proof without knowing the correct witness.

Together, these elements and checks allow zk-SNARK verification to be completed in constant time, using only a small number of elliptic curve pairings, without ever revealing the witness.

\subsection{Recurrence Constraints in STARKs}
\label{app:recurrence-constraints}

In the implementation, the Fibonacci-like recurrence 
\[
a_{n+2} = a_{n+1}^2 + a_n^2
\]
is encoded algebraically on the interpolated trace polynomial $P(x)$. 
Specifically, the protocol checks whether
\[
P(g^2 x) - P(g x)^2 - P(x)^2 = 0
\]
for all $x$ in the subgroup $G$, where $g$ is the generator of $G$. 

Alongside this recurrence relation, two boundary constraints are enforced: 
\[
P(1) = a_0 = 1, 
\quad P(g^{1022}) = a_{1022} = 2338775057.
\]

These constraints are then combined into a single composition polynomial $C(x)$, using random linear coefficients derived from the Fiat--Shamir transcript. This composition ensures that all recurrence and boundary conditions are compactly verified within the low-degree testing framework of the STARK protocol.

\end{document}